\documentclass[aps,pra,twocolumn,superscriptaddress,english,floatfix]{revtex4-2}
\usepackage{graphicx}
\usepackage{float}
\usepackage{physics}
\usepackage{cancel}
\usepackage{overpic}
\usepackage{enumerate}
\usepackage{hyperref}
\usepackage{textcomp}
\usepackage{amsmath}
\usepackage{amssymb}
\usepackage{pgf}
\usepackage{soul}
\usepackage[normalem]{ulem}
\usepackage{lipsum}
\usepackage{comment}
\usepackage{mathrsfs}
\usepackage{standalone}
\usepackage[english]{babel}
\usepackage{bm}
\usepackage{tikz}
\usetikzlibrary{shapes.geometric}
\usetikzlibrary{decorations.markings}
\usetikzlibrary{decorations.pathmorphing,patterns}
\usetikzlibrary{decorations}
\usetikzlibrary{calc}
\usetikzlibrary{intersections}
\usetikzlibrary{arrows}
\usetikzlibrary{matrix}
\usetikzlibrary{positioning}
\usetikzlibrary{external}
\newcommand{\btheta}{\boldsymbol{\theta}}

\newcommand{\eff}{\mathrm{eff}}

\newcommand{\bfv}{\mathbf{v}}

\newcommand{\D}{\mathcal D}
\newcommand{\N}{\mathcal N}
\newcommand{\Heaviside}{\Theta}
\newcommand{\OP}{\psi}
\newcommand{\La}{\mathscr{L}}
\newcommand{\force}{F}
\newcommand{\ifunc}{\mathscr F}

\graphicspath{{./}{./s/}}

\begin{document}

\title{Quantum Effects on the Synchronization Dynamics of the Kuramoto Model}
\author{Anna Delmonte}
\affiliation{SISSA, Via Bonomea 265, I-34136 Trieste, Italy}
\author{Alessandro Romito}
\affiliation{Department of Physics, Lancaster University, Lancaster LA1 4YB, United Kingdom}
\author{Giuseppe E. Santoro}
\affiliation{SISSA, Via Bonomea 265, I-34136 Trieste, Italy}
\affiliation{The Abdus Salam International Center for Theoretical Physics, Strada Costiera 11, 34151 Trieste, Italy}
\affiliation{CNR-IOM, Consiglio Nazionale delle Ricerche - Istituto Officina dei Materiali, c/o SISSA Via Bonomea 265, 34136 Trieste, Italy}
\author{Rosario Fazio}
\affiliation{The Abdus Salam International Center for Theoretical Physics, Strada Costiera 11, 34151 Trieste, Italy}
\affiliation{Dipartimento di Fisica ``E. Pancini", Universit\`a di Napoli ``Federico II'', Monte S. Angelo, I-80126 Napoli, Italy}

\begin{abstract}

The Kuramoto model serves as a paradigm for describing spontaneous synchronization in a system of classical interacting rotors. In this study, we extend this model to the quantum domain by coupling quantum interacting rotors to external baths following the Caldeira-Leggett approach. Studying the mean-field model in the overdamped limit using Feynman-Vernon theory, we show how quantum mechanics modifies the phase diagram. Specifically, we demonstrate that quantum fluctuations hinder the emergence of synchronization, albeit not entirely suppressing it. We examine the phase transition into the synchronized phase at various temperatures, revealing that classical results are recovered at high temperatures while a quantum phase transition occurs at zero temperature. Additionally, we derive an analytical expression for the critical coupling, highlighting its dependence on the model parameters, and examine the differences between classical and quantum behavior.
\end{abstract}

\maketitle

\section{Introduction}
Synchronization is an emergent collective phenomenon that can be observed in various physical systems, such as pendula \cite{Huygens:letters}, fireflies \cite{Morse:article,Buck:article}, and neurons \cite{Gray:article}. In classical mechanics, synchronization can occur when two or more oscillators interact with each other through a common coupling \cite{Gupta:book,pikovsky:book}.
Classical synchronization has been witnessed in systems that can operate in the quantum regime \cite{Wang:article,Weisenfeld:article,Matheny:article}, which is nowadays accessible to experiments due to the recent advancements in the field of quantum technologies. For example, optomechanical devices \cite{Machado:article,Lemonde:article,Weiss:article} have allowed the coupling between light and mechanical motion to be controlled, leading to the possibility of implementing non-linear dynamics that can result in a synchronized motion.

These perspectives, with their possible applications in quantum technologies, have also posed a number of new questions on how to characterize and quantify synchronization in quantum systems. Synchronization becomes even more intriguing, as it has to deal with quantum fluctuations and entanglement. Furthermore, it may be a useful resource for quantum technological applications \cite{Agnesi:article,Calderaro:article}, for example in quantum thermal machines \cite{Jaseem2:article,Solanki:article,murtadho2:arxiv}. An intense theoretical activity aimed at quantifying synchronization in quantum systems from continuous variables \cite{Arosh:article,Sonar:article,Chia:article,Walter:article,Walter2:article,Lee:article,Lee2:article,Lorch:article,Zhirov3:article,Cabot2:article} to discrete degrees of freedom \cite{Roulet:article,Roulet2:article,Koppenhofer:article,Krithika:article,Solanki:article,Zhirov1:article,Zhirov2:article,Giorgi:article,Buca:article,Bellomo:article,Fistul:article,Cattaneo:article}. Different measures of synchronization have been introduced, ranging from phase-space or correlation quantities \cite{Mari:article,Jaseem:article,Kato:article,Kato2:article,Hush:article,Cabot:article} to information-theoretical approaches \cite{Jaseem:article,Eneriz:article,Galve:book}.

This large body of work, however, did not address a seemingly natural question: how to extend a paradigmatic model of classical synchronization, the Kuramoto model \cite{kuramoto}, to study synchronization in the presence of quantum fluctuations \cite{kuramoto}. 
This is the starting point for our work. The classical model describes the behaviour of interacting rotors with a non-linear dynamics, and exhibits a phase transition from a dynamically disordered phase, to an ordered one characterized by phase locking. Generalizations of the model \cite{SakaguchiKuramoto1:article,SakaguchiKuramoto2:article,Acbron:review,Gherardini:article,Strogatz:article,Olmi:article,Daido:article} have allowed to explore and enrich the phase diagram by studying also the effects of noise, inertia, disorder and long-range interactions on the emergence of synchronization.
Despite efforts to study and understand the emergence of collective behavior in a semiclassical regime, where quantum fluctuations become relevant and modify the system's dynamics \cite{Zueco:article}, a systematic analysis of spontaneous synchronization in the fully quantum regime is still lacking. 
Can quantum synchronization emerge in a low-temperature regime or do quantum fluctuations dominate the system's behavior,  preventing spontaneous synchronization?

In this paper, we address this problem by exploring whether the Kuramoto model can be 
extended to the quantum regime. 
We study  the dynamics of the model from high to low temperature and show that synchronization survives quantum fluctuations and a quantum phase transition is still present in the zero-temperature limit.

The rest of the article is organized as follows:
in Sec.~\ref{sec:classic} we describe the celebrated Kuramoto model with a path integral formalism. We also present its phase diagram and the most relevant results, focusing in particular on the generalized massive model.
In Sec.~\ref{sec:quantum} we propose a new quantum model, based on the classical model we discuss in \ref{sec:classic}. The limits in which the model is studied are discussed, and the order parameter to detect quantum synchronization is defined. In this section we also show that in the high-temperature regime, our model reproduces correctly the classical one.    
In Sec.~\ref{sec:selfconeq} we introduce the self-consistent equation to determine the order parameter. The self-consistent equation allows us to study the phase diagram of the quantum model in the overdamped regime, and in particular to determine analytically the critical coupling above which the system enters a synchronized phase. The results of this analysis are reported in Sec.~\ref{sec:results}. The last section, Sec. \ref{sec:conclusions}, we present some conclusions that can be drawn from our study.

\section{Classical Kuramoto model} \label{sec:classic}
\begin{figure}[h]
{\includegraphics[width=.3\textwidth]{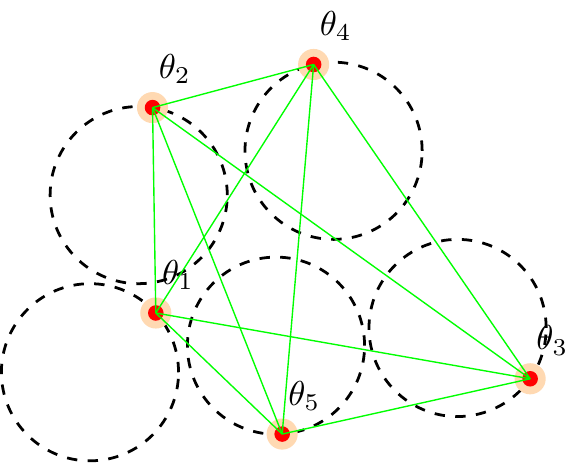}}
\caption{Representation of the classical Kuramoto model for $N=5$. A set of rotors (red dots on dashed circles) evolve with independent frequencies. Their mutual interactions (green lines) induce a phase transition to  synchronised dynamics. }
\label{fig4}
\end{figure}

The Kuramoto model describes the behaviour of $N$ interacting planar rotors.
It exhibits two phases: an incoherent phase characterized by the rotors moving independently, and a synchronized phase in which the system behaves collectively. The mechanism underlying in the synchronization process is \textit{phase locking}, it causes the emergence of a fixed relation between the phases of the rotors.

The state of each rotor is characterised by a phase $\theta_i$, an angular velocity $v_i$, with $i=1,...,N$. The evolution of this state is determined by a frequency $\omega_i$, and damping $\gamma$. The characteristic frequencies $\omega_i$ are independent of each other and, throughout the paper, will be drawn from an even, unimodal frequency distribution $g(\omega)$, with average $\overline\omega=\expval{\omega}_{g(\omega)}=0$ and variance $\sigma^2$.

We consider here the massive version of this model described by the following set of Langevin equations:
\begin{equation}\label{eq_Lang}
    \begin{cases}
    &\dot\theta_i=v_i \vspace{3mm} \\
    &m\dot v_i+m\gamma v_i=\force[\btheta;\omega_i]+\xi_i
    \end{cases}
    \vspace{1mm}
\end{equation}
Here, $\btheta=\left(\theta_{1},...,\theta_{N}\right)$ and
\begin{equation}
    \force[\btheta;\omega_i]=\omega_i-\frac{J}{N}\sum\limits_{j=1}^N\sin(\theta_i-\theta_j) \;.
\end{equation}
The noise in the Langevin equation is a Gaussian stochastic process with $\expval{\xi_i(t)}=0$, $\expval{\xi_i(t)\xi_j(t')}=2D\delta_{i,j}\delta(t-t')$.
The initial conditions $\btheta(0)=\btheta_0$,  $\bfv(0)=\bfv_0$ are drawn independently for every rotor from a distribution $\rho(\theta_0,v_0)$.

In the massless limit ($m\gamma=\rm{const}$, $m\to0$) the Langevin equation reduces to the Kuramoto-Sakaguchi model \cite{Sakaguchi:article} 
\begin{equation}
\dot\theta_i=\force[\btheta,\omega_i]+\xi_i \hspace{10mm} 
\forall i=1,..,N \;.
\end{equation}

The synchronized behaviour is signaled by a non-zero value, in the stationary state, of the modulus of the complex order parameter $\OP(t)$ defined as
\begin{equation} \label{eq_Op}
\OP(t)=r(t)\, e^{i\varphi(t)} = \frac{1}{N}\sum_{j=1}^Ne^{i\theta_j(t)} \;.
\end{equation}
The modulus $r$ of the order parameter $\psi$ is bounded to the interval $r(t)\in[0,1]$. It efficiently detects synchronization since it averages to zero if the rotors evolve incoherently. The phase $\varphi$ corresponds to the phase of the collective motion of the rotors.

In the thermodynamic limit, the definition of the order parameter is regarded as an average over the frequency distribution, the noise distribution and the distribution of the initial conditions, namely
\begin{equation}\label{eq_OpAverage}
    \OP(t)=\int_{-\infty}^{\infty}\!\! d\omega \, g(\omega)
    \expval{e^{i\theta(t;\omega,\xi)}}_{\xi,\theta_0,v_0} \;,
\end{equation}
where the phase $\theta(t;\omega,\xi)$ satisfies the Langevin equation 
\eqref{eq_Lang} with 
$\force[\btheta;\omega]=\force[\theta;\omega,\psi]=\omega-Jr\sin(\theta-\varphi)$. 
Notice that, to formally decouple the evolution of the rotors, we have used
$\frac{J}{N}\sum_{j=1}^N\cos(\theta_i-\theta_j)=\frac{J}{2}e^{i\theta_i}\left(\frac{1}{N}\sum_{j=1}^Ne^{-i\theta_j}\right)+\rm{c.c.}=Jr\cos(\theta_i-\varphi).$ This results in \eqref{eq_OpAverage} becoming a self-consistent equation for the order parameter.

For further convenience it is useful to express the average that defines the order parameter in a path integral form \cite{Acbron:review}.
Discretizing {\em à la Ito}, one can express \eqref{eq_OpAverage} as (see Appendix \ref{app:pi_class} for details):
\begin{equation}
\psi(t)=\int_{0}^{2\pi} \! d\theta \, e^{i\theta}
\int_{-\infty}^{\infty} \! dv 
\int_{-\infty}^{\infty} \! d\omega \, g(\omega) \, \rho(\theta,v,t;\omega,\psi) \;.    
\end{equation}.
\begin{widetext}
Here $\rho(\theta,v,t;\omega,\psi)$ is a probability distribution that quantifies the probability for the $i^{th}$ rotor to have phase and angular velocity $(\theta,v)$ at time $t$, and can be expressed as \cite{Cugliandolo:article,Das:article,Tauber:book}, \cite{Zinn-Justin:book}[Chap.4].
\begin{eqnarray}\label{eq_Jclass}
\rho(\theta,v,t;\omega,\psi) = \N 
\int_0^{2\pi} \! d\theta_0 
\int_{-\infty}^{\infty} \! d v_0 
\int_{\theta(0)=\theta_0}^{\theta(t)=\theta} \! \D\theta
\int_{v(0)=v_0}^{v(t)=v} \! \D v
\, \delta(\dot\theta(\tau)-v(\tau)) \, 
e^{iS_{cl}[\theta(\tau),v(\tau)]} \, \rho(\theta_0,v_0),
\end{eqnarray}
where the classical action is given by:
\begin{equation}
S_{cl} = \frac{i}{4D} \int_0^t dt'
\Big(m\dot v(t')+m\gamma v(t')-\force[\theta(t');\omega,\psi(t')]\Big)^2 \;.
\end{equation}
\end{widetext}

Solving the self-consistent equation for the order parameter allows to gain knowledge about the phase transition. The behavior of the system is determined by the interplay between the coupling strength $J$, the width of the frequency distribution, $\sigma$, and of the noise distribution, $D$. In general, the phase transition for the model described by Eq.~\eqref{eq_Lang} is first-order, but it becomes a continuous phase transition in the overdamped limit \cite{Gupta:book,Acbron:review}.

An interesting case is the massless model, for which the critical coupling is known analytically to be
\begin{equation}
J_C^{cl}=2\Big(\int_{\infty}^{\infty} \! d\omega \, g(\omega) \, \frac{D}{D^2+\omega^2}\Big)^{-1} \;.
\end{equation}
From this formula, the effects of noise and the width of the frequency distribution are evident. Noise hinders the phase-locking mechanism, and so does increasing the variance of the frequency distribution. In the case of the noiseless model, the critical coupling becomes simply $J_C^{cl}=\frac{2}{\pi g(0)}$. 

\section{Quantum Kuramoto model}\label{sec:quantum}
\begin{figure}[h]
{\includegraphics[width=.35\textwidth]{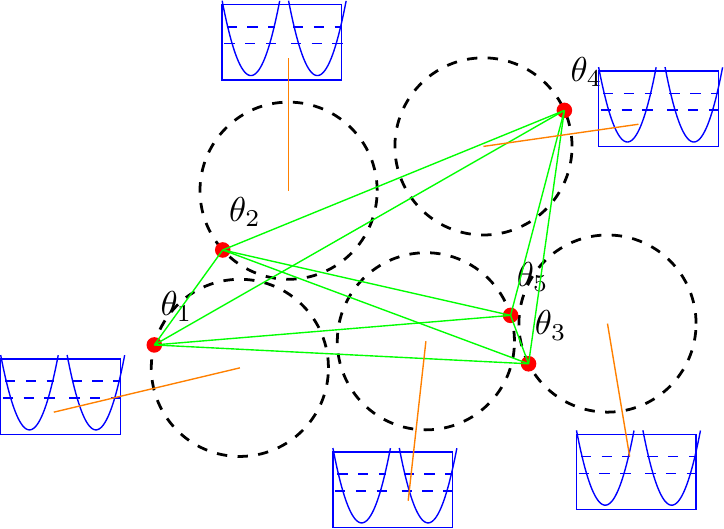}}
\caption{Representation of the quantum Kuramoto model. Same as  Fig.1, with the rotors as quantum systems. The coupling to independent identical quantum baths is explicitly shown as an orange line connecting to a set of harmonic oscillators (blue boxes)}
\label{fig5}
\end{figure}

Our goal is now to construct a model that reproduces the classical one in the high-temperature limit.
Since the classical Kuramoto-Sakaguchi model is characterized by noise and dissipation, it cannot be obtained as a classical limit of a quantum Hamiltonian model.
Thus, we introduce dissipation in the quantum regime via a Caldeira-Leggett \cite{CaldeiraLeggett:article} model made of $N$ interacting rotors, each one linearly coupled to a different and independent bath of harmonic oscillators. The baths are assumed to be identical. 

The Lagrangian describing this model is
\begin{equation}
    \La_{\rm{TOT}} = \La_S + \La_B + \La_{SB} \;.
\end{equation}
$\La_S$ is the Lagrangian of the system of rotors
\begin{equation}\label{eq_sys}
    \La_S = \sum_{i=1}^N \left[\frac{m\dot\theta_i^2}{2} + \omega_i\theta_i + \frac{J}{2N}\sum_{j\not= i}\cos{\left(\theta_i-\theta_j\right)}\right] \;.
\end{equation}
The characteristic frequencies $\omega_i$ are once again drawn from a distribution $g(\omega)$ having the same characteristics as in the classical case.
$\La_B$ is the baths' Lagrangian:
\begin{equation}\label{eq_bath}
    \La_B = \sum_{i=1}^N\La_{B_i} = \sum_{i=1}^N\sum_{j_i=1}^{M}\left[\frac{M\dot x_{j_i}^2}{2}-\frac{1}{2}M\Omega_{j_i}^{2}x_{j_i}^2\right] \;.
\end{equation}
$\La_{SB}$ is the interaction Lagrangian
\begin{equation}\label{eq_int}
    \La_{SB}=\sum_{i=1}^N\La_{SB_i}=C\sum_{i=1}^N\theta_i\sum_{j_i=1}^{M}x_{j_i} \;.
\end{equation}

In order to decouple the equations, as in the classical case, it is convenient to work in the thermodynamic limit with a mean-field model.
We assume $\frac{1}{N}\sum_{j=1}^Ne^{i\theta_j} = \expval{\frac{1}{N}\sum_{j=1}^Ne^{i\theta_j}}+\delta\psi$ 
with $\delta\psi$ infinitesimal and define
$\psi=re^{i\varphi}=\expval{\frac{1}{N}\sum_{j=1}^Ne^{i\theta_j}}$ where the quantum averages are taken over the rotors' reduced density matrix.
The mean-field Lagrangian $\La$ becomes up to first order in $\delta\psi$
\begin{equation}\label{eq_Lagrangian}
\La=\sum_{i=1}^N\Big(\La_{S_i}+\La_{B_i}+\La_{SB_i}\Big) \;,
\end{equation}
with $\La_{S_i}=\frac{m\dot\theta_i^2}{2} + \omega_i\theta_i + Jr\cos(\theta_i-\varphi)$. 
Notice that the mean-field model is described by a Lagrangian decoupled in a sum of terms depending only on the $i^{th}$ rotor.
From now on for brevity 
\begin{equation}\label{eq_pot}
    V[\theta]=-\omega\theta-Jr\cos\left(\theta-\varphi\right) \;.
\end{equation}

In order to reproduce the friction term in Eq. \eqref{eq_Lang} in the classical limit, The Caldeira-Leggett's model requires an Ohmic bath \cite{CaldeiraLeggett:article}. 
We therefore demand that, for each and every bath, the distribution of frequencies for the collection of harmonic oscillators is
\begin{equation}\label{eq_ohmicdist}
    \sum_{i=1}^M\frac{C^2}{2M\Omega_i}\,\delta(\Omega_i-\nu)\,\xrightarrow[M\to\infty]{}\,   \frac{m\gamma \nu}{\pi} \Heaviside(\omega_c-\nu) \;,
\end{equation}
where $\Heaviside(\cdot)$ is the Heaviside fucntion and $\omega_c$ is a cutoff for the frequencies of the baths' harmonic oscillators.

A few comments about the definition domain of the phases $\theta_i$ are in order. Two choices are possible: the phases of the rotors can be defined over the circle, i.e. $\theta_i\in\left[0,2\pi\right]$, or they can be defined over the line $\theta_i\in\mathbb R$. 
The difference lies in identifying or not the position $\theta$ with $\theta+2n\pi$, $n\in\mathbb Z$. 
The Langevin equation that describes the classical model in 
Eq.~ \eqref{eq_Lang} is not dependent on the choice of the phases' domain since it is inherently $2\pi$ periodic. 
For the quantum model we define $\theta_i\in\mathbb R$. 
The tilted potential term $\omega\theta$ and the linear coupling with the bath are compatible with this choice~\cite{Weiss:book}[Chap.2]. 
We note that with this choice of phase domain, the Lagrangian in Eq.~\eqref{eq_sys} can be regarded as the Lagrangian of a resistively shunted Josephson Junction \cite{Zwerger:article}.

In order to understand if the system described by the mean-field Lagrangian \eqref{eq_Lagrangian} can sustain synchronization, an order parameter should be defined. In analogy with the classical case, we define
\begin{equation}
\psi(t)=r(t) \, e^{i\varphi(t)} = 
\frac{1}{N} \sum_{j=1}^N\Tr{e^{i\theta_j} \, \rho_S(t)}
\end{equation}
where $\rho_S(t)$ is the reduced density matrix of the system of rotors evolved to time $t$ after tracing out the baths' degrees of freedoms. Notice that, once again, in the mean-field approximation this is a self-consistent equation for the order parameter, since the density matrix evolves with a Lagrangian dependent on $\psi$.

The initial state of the evolution is chosen to be separable in the rotors; the mean-field approximation and the choice of independent baths allow to maintain the density matrix separable in the rotors at any time. 
For this reason, from now on, the discussion will focus only on the evolution of the density matrix $\rho_{SB}^{(i)}$ of a single rotor and its own bath. 
The initial density matrix for the $i^{th}$ rotor and the bath is also assumed to be separable: $\rho_{SB}^{(i)}=\rho_i\otimes\rho_{B}$.

\begin{widetext}
The evolution of the reduced density matrix of a single rotor can be obtained through Feynman-Vernon method \cite{FeynmanVernon:article,FeynmanHibbs:book}, appropriate to treat quantum systems with a classical limit given by a stochastic equation of motion \cite{Schmid:article,Grabert:article,Fisher:article,HakimAmbegaokar:article}. Applying the Feynman-Vernon method along the lines of Ref. \cite{CaldeiraLeggett:article},
the density matrix element $\rho_i(\theta_1,\theta_2)=\langle \theta_2| \hat{\rho}_i | \theta_1\rangle$ is given at time $t$ by
\begin{eqnarray}\label{eq_rhoev}
\rho_i(\theta_1,\theta_2,t)=
\int_{-\infty}^{\infty} \! d\theta'_1
\int_{-\infty}^{\infty} \! d\theta'_2 \int_{\theta_1(0)=\theta'_1}^{\theta_1(t)=\theta_1} \D\theta_1 \, \int_{\theta_2(0)=\theta'_2}^{\theta_2(t)=\theta_2} \D\theta_2 \,
e^{\frac{i}{\hbar}\left(S_0[\theta_1]-S_0[\theta_2]\right)} \,
\ifunc[\theta_1,\theta_2] \, \rho_{i}(\theta'_1,\theta'_2,0)
\end{eqnarray}
where $S_0=\int_0^tdt'\La_{S_i}(t')$ for the $i^{th}$ rotor, and $\ifunc[\theta_1,\theta_2]$ is the Feynman-Vernon's influence functional that accounts for the effects of the interaction with the bath.

We can regard the previous equation as an evolution of the density matrix due to an effective action 
$S_{\eff}[\theta_1,\theta_2]=S_0[\theta_1]-S_0[\theta_2]-i\hbar\ln{\ifunc[\theta_1,\theta_2]}$. 
Switching to the more convenient variables 
$\theta_+=(\theta_1+\theta_2)/2$, $\theta_-=\theta_1-\theta_2$:
\begin{eqnarray}\label{eq_qeffeacti}
S_{\eff}[\theta_+,\theta_-] = \int_0^t \! dt' \,
\Big( m\dot\theta_{+} \dot\theta_- 
- \sum_{q=\pm1} (-1)^{q} V[\theta_+ + q{\textstyle\frac{\theta_-}{2}}] - 
 m\gamma\theta_{-} \dot{\theta}_{+} 
 + \frac{iD}{\hbar} \int_0^t\! dt''\,\theta_{-}(t') K(t'-t'') \theta_{-}(t'')
\Big) \;.
\end{eqnarray}
In the previous equation $D=m\gamma k_BT$, and $K(t)$ is given by the Fourier transform $K(t)=\int_{-\omega_c}^{\omega_c}\frac{d\nu}{2\pi}\mathcal K(\nu)e^{-i\nu t}$, with 
\begin{equation}
\mathcal K(\nu)=\frac{\hbar\nu}{2k_BT}\coth{\left(\frac{\hbar\nu}{2k_BT}\right)} \;.
\end{equation}
%
The temperature $T$ is set by the bath, and the signatures of its interaction with the rotor are the friction term $m\gamma\theta_-\dot\theta_+$ and the imaginary term containing the memory kernel $K(t)$. These terms are given, respectively, by the imaginary and real part of the bath's correlation function ($\hbar\alpha(t'-t'')$ in the notation of Ref.~\cite{CaldeiraLeggett:article}). 
In  Eq. \eqref{eq_qeffeacti}, we have neglected a Lamb-shift term in the energy originating from the influence functional, which does not affect the dynamics of the model \cite{Schmid:article}.

Before proceeding with the calculation of the order parameter, it is worth noticing how the classical dynamics is recovered in the high temperature limit of the quantum model. 
In the infinite temperature limit, the memory kernel becomes $K(t)\xrightarrow[]{T\to\infty}\delta(t)$ and the imaginary damping term prevents $\theta_-$ to vary \cite{Schmid:article}, \cite{Weiss:book}[Chap.5]. 
This means that, starting from a ``classical'' diagonal state with $\theta_-(0)=0$, the density matrix will always remain diagonal (the off-diagonal term are exponentially suppressed). 
Moreover, expanding up to first order $\theta_-(t)$, the evolution becomes
\begin{eqnarray}\label{eq_classexp}
\rho_i(\theta_+,\theta_-,t)
= \int_{-\infty}^{\infty} \! d\theta'_{+}
\int_{\theta_+(0)=\theta'_+}^{\theta_+(t)=\theta_+} \!\! \D\theta_+
\int_{\theta_-(0)=\theta'_-=0}^{\theta_-(t)=0} \!\! \D\theta_-
\, e^{\frac{i}{\hbar} S_{\eff}[\theta_+,\theta_-]} \, \rho_{i}(\theta'_+,\theta'_-=0,0) \;,
\end{eqnarray}
with the effective action being
\begin{eqnarray}
S_{\eff}[\theta_+,\theta_-] =
\int_0^t \! dt'\Bigg(\frac{iD}{\hbar}\theta_-^2(t') 
- \theta_-(t') \Big(m\ddot\theta_+ + m\gamma\dot\theta_+ - 
\omega + Jr\sin(\theta_+-\varphi)\Big) \Bigg) \;,
\end{eqnarray}
where we have used the fact that, at first order in $\theta_-$, $V[\theta] \approx  \theta_-\force[\theta_+]$ . 
 With the change of variables $\frac{\theta_-(\tau)}{\hbar}\to\eta(\tau)$, 
one recovers the classical effective action for the stochastic process \eqref{eq_Lang} showed in Eq.~\eqref{eq_classaction} after the integration of the $\delta$-function  in the angular velocity. Thus, the quantum model reproduces correctly the massive Kuramoto-Sakaguchi model in the infinite temperature limit.
\end{widetext}

\subsection{The self-consistent equation for the order parameter}\label{sec:selfconeq}

In order to obtain a self-consitent equation for the order parameter, we start noticing that the reduced density matrix of the rotors at time $t$ is given by $\rho_S(t)=\bigotimes_{i=1}^N\rho_i(t)$.
Eq. (17) then takes the form
\begin{eqnarray}
    \psi(t)&=&\frac{1}{N}\sum_{j=1}^N\,\Tr{e^{i\theta_j}\,\rho_j(t)}\bigotimes_{k\not =j, k=1}^{N}\Tr{\rho_k}\\&=&\frac{1}{N}\sum_{j=1}^N\Tr{e^{i\theta_j}\,\rho_j(t)}\nonumber\;.
\end{eqnarray}

In the thermodynamic limit the previous expression corresponds to an average over the frequency distribution $g(\omega)$. Denoting as $\rho(t)$ the density matrix of a single rotor we have: 
\begin{equation}\label{eq_selfcon}
\psi(t) = \int_{-\infty}^{\infty} \! d\omega \, g(\omega) 
\int_{-\infty}^{\infty} \! d\theta_+ 
\, e^{i\theta_+} \, \rho(\theta_+,\theta_-=0,t) \;.
\end{equation}
To get a more explicit expression of the self-consistent equation for the order parameter, the path integration in Eq.~\eqref{eq_rhoev} should be performed. 
The specific form of the potential \eqref{eq_pot} appearing in the effective action does not allow for a general calculation of the path integral. However, the knowledge of the behaviour of the classical model helps: in the overdamped limit the classical phase transition to the synchronized phase is of second-order.
We expect that the quantum phase transition in the overdamped limit is second order too. If this is the case, assuming $J\sim J_C$, we can perform a perturbative expansion in $r$ for $r\sim0$ of \eqref{eq_selfcon} to gain insight onto the quantum dynamics. 
Thus, we will hereafter focus explicitly on the overdamped regime $\frac{m\gamma}{\hbar}\gg1$ of the model.

The perturbative expansion of terms in the evolution of the density matrix in Eqs. \eqref{eq_rhoev},\eqref{eq_qeffeacti} involves only the approximation
\begin{widetext}
\begin{eqnarray}\label{eq_expansion}
\exp{\frac{-iJr}{\hbar} \int_0^t dt' \sum_{q=\pm} 
q \cos(\theta_+(t') - q{\textstyle \frac{\theta_-(t')}{2}} - \varphi(t'))} 
\sim 1 
&-&\frac{iJr}{\hbar}\int_0^t ds \cos(\theta_+(s)-{\textstyle\frac{\theta_-(s)}{2}} -\varphi(s)) \nonumber \\
&+&\frac{iJr}{\hbar}\int_0^t ds \cos(\theta_+(s)+{\textstyle\frac{\theta_-(s)}{2}}-\varphi(s)) \;. \nonumber
\end{eqnarray}
\end{widetext}
The zeroth order of the expansion of the density matrix \eqref{eq_rhoev} gives an evolution according to a Lagrangian with potential $V[\theta]=-\omega\theta$. 
This Lagrangian does not contain terms that can synchronize the rotors, hence it does not contribute to the self-consistent equation for the order parameter. 
The first order expansion, together with the {\em Ansatz} $\varphi(t)=0$ and $r\sim 0$ constant, yields, for $t \to \infty$ 
\begin{equation}\label{eq_1ord}
r = r J_C \lim_{t\to\infty}\, \int_{-\infty}^{\infty} \!\! d\omega \, g(\omega) \int_{-\infty}^{\infty} \! d\theta_+ e^{i\theta_+}\rho'(\theta_+,\theta_-,t)
\end{equation}
where 
\begin{widetext}
\begin{eqnarray}
\rho'(\theta_+,0,t) &=& \frac{-i}{2\hbar}\sum_{c,c'=\pm1} c' \int_0^t ds \int_{-\infty}^{\infty} d\theta'_{+} \int_{-\infty}^{\infty} d\theta'_{-} \int_{\theta_+(0)=\theta'_+}^{\theta_+(t)=\theta_+} \D\theta_+ \int_{\theta_-(0)=\theta'_-}^{\theta_-(t)=0} \D\theta_- 
\nonumber \\
&& \hspace{40mm} \exp{\frac{i}{\hbar} S'_{\eff}[\theta_+,\theta_-;s,t]_{c,c'}} \, \rho(\theta'_+,\theta'_-,0) \;,
\end{eqnarray}
and 
\begin{eqnarray}\label{eq_effact}
S'_{\eff}[\theta_+,\theta_-;s,t]_{c,c'} &=& S'_{Re}[\theta_+,\theta_-;s,t]_{c,c'} + i \, S'_{Im}[\theta_-;s,t]_{c,c'} \nonumber \\ 
&=& 
\int_0^t dt' \biggl\{m \, \dot\theta_{+}\dot\theta_- - m\gamma \, \theta_{-}\dot{\theta}_+ + \omega \, \theta_- \nonumber \\ 
&& \hspace{10mm} + i\hbar c \, \theta_+(t')\delta(t'-s) 
-\frac{i\hbar cc'}{2} \, \theta_-(t')\delta(t'-s) + \frac{iD}{\hbar}
\int_0^t dt'' \theta_{-}(t') K(t'-t'') \theta_{-}(t'')\biggr\} \;. \hspace{2mm}       
\end{eqnarray}
The details of the derivation of Eqs. \eqref{eq_effact} can be found in Appendix \eqref{app:firstorder}.
\end{widetext}

If Eq.~\eqref{eq_1ord} admits a solution for $r \neq 0$,  then the model admits a phase transition to a synchronized state in the overdamped limit, and the resulting $J_C$ gives the value of the critical coupling for the phase transition. Notice that the first order expansion does not contain information about the order of the phase transition, that could only be understood through a third order expansion. The goal is now to find the critical coupling and to study its dependence on the parameters that characterize the system: $m,\gamma,k_BT$, and the variance $\sigma$ of the even unimodal frequency distribution $g(\omega)$.   

The expansion to first order in the self-consistent equation has produced a Gaussian path integral that can now be performed \cite{HakimAmbegaokar:article,Fisher:article}.
The calculation can be performed via a decomposition of the effective action \eqref{eq_effact} in its real and imaginary part 
$S'_{\eff}=S'_{Re}[\theta_+,\theta_-]+iS'_{Im}[\theta_-]$, 
as can be seen from the previous equation.

\begin{widetext}
The calculation, reported in Appendix \ref{app:firstorder}, produces the following result for the first order expansion of the self-consistent equation:
\begin{eqnarray}\label{eq_scintermediate}
r &=& r J_C \lim_{t\to\infty} \int_{-\infty}^{\infty} \! d\omega \, g(\omega) \int_{-\infty}^{\infty} \! d\theta_+ \, e^{i\theta_+}
\left(\frac{-i}{2\hbar}\right) \, \frac{m\gamma}{2\pi\hbar} 
\sum_{c,c'=\pm 1} c' \int_0^t ds \int_{-\infty}^{\infty} d\theta'_{+}
\int_{-\infty}^{\infty} \! d\theta'_{-} \, \rho(\theta_+',\theta'_-,0) \nonumber \\ 
&& \hspace{20mm} \text{exp} \biggl\{- \, \frac{1}{\hbar} S'_{Im}[\Tilde\theta_-;s,t]_{c,c'} -
\frac{i}{\hbar}m \, \theta_-'\dot{\Tilde\theta}_+(0) + ic \, \Tilde\theta_+(s)\biggr\} 
\end{eqnarray}
\end{widetext}
where $\Tilde\theta_\pm$, are the solutions of the equations
\begin{equation}\label{eq_sp_1_2}
\left\{ 
\begin{array}{rcl}
\ddot{\Tilde\theta}_-(t') &-& \gamma\dot{\Tilde\theta}_-(t')-\frac{\hbar c}{m}\delta(t'-s)=0 \vspace{3mm} \\
\ddot{\Tilde\theta}_+(t') &+& \gamma\dot{\Tilde\theta}_+(t')-\frac{\omega}{m}+\frac{\hbar cc'}{2m}\delta(t'-s)=0
\end{array} \right. \;.
\end{equation}
The explicit solution of these equations is reported in appendix \ref{app:sp}.

The term $c=1$ in the sum does not contribute because it gets completely damped by the imaginary part of the action. From now on we will consider only the term $c=-1$. The expression for $\dot{\Tilde\theta}_+(0)$ and $\Tilde\theta_+(s)$ can be deduced from the result in appendix \ref{app:sp}. It is convenient to proceed by calculating the integration over $\theta_+$ first. Isolating the integration and all the terms of \eqref{eq_scintermediate} involving $\theta_+$, we find
\begin{equation}
\frac{m\gamma}{2\pi \hbar}\int _{-\infty}^{\infty } \!\! d\theta_{+}e^{i\theta_{+}\left(-\frac{m\gamma}{\hbar}\theta'_{-} - \, e^{-\gamma s}) \right)} = 
\delta\left(\theta_{-}'-{\textstyle\frac{\hbar}{m\gamma}} 
e^{-\gamma s}\right) \;.
\end{equation}
This constraints  the initial value of $\theta_-$ as determined by the delta function. In the massless limit ($\gamma\to\infty$) the value that the delta function selects is $\theta_-'=0$, 
in the general overdamped limit $\frac{m\gamma}{\hbar}\gg1$ ($m$ and $\gamma$ finite) we select $\theta'_-\sim 0$. 
Performing now the integration over $\theta'_-$ to eliminate the delta function and denoting 
$\Tilde\theta_-(t')\bigl|_{\theta'_- = \frac{\hbar}{m\gamma}e^{-\gamma s}}$ as $\theta_-^*(t';s,t)$, the self-consistent equation becomes
\begin{widetext}
\begin{eqnarray}
r = \frac{-iJ_Cr}{2\hbar}\lim_{t\to\infty}\int_{-\infty}^{\infty} \! d\omega \, g(\omega) \, \int_0^t \! ds \,
\sum_{c'=\pm1} c' \exp{ic'\frac{\hbar}{2m\gamma}
-i \, \frac{\omega(t-s)}{m\gamma}- \, \frac{1}{\hbar}S'_{Im}[\theta_-^*;s,t]}
\end{eqnarray}
where we have used the fact that in the overdamped limit 
\[ 
\int_{-\infty}^{\infty} \! d\theta_+' \, \rho(\theta_+',\frac{\hbar}{m\gamma} \, e^{-\gamma s})
\sim \int_{-\infty}^{\infty} \! d\theta_+' \, \rho(\theta'_+,0) = \Tr{\rho(t=0)}=1 \;. 
\]
\end{widetext}
This approximation becomes exact in the massless limit. Also notice that the sign of the sine, $\sin{\left(\frac{\hbar}{2m\gamma}\right)}\sim\frac{\hbar}{2m\gamma}$, is always defined (and positive) in the overdamped limit, and that the expression does not depend on the initial state of the system.

We conclude that a non-trivial solution to the self-consistent equation exists (if the time limit exists), thus a phase transition happens at $J_C$.

\section{Results} \label{sec:results}
\vspace{-0.52cm}

From the analysis in Section \ref{sec:selfconeq}, and Eq. (31) in particular, it follows that  the quantum Kuramoto model in the overdamped limit admits a transition to a synchronized phase with a critical coupling  given by
\begin{widetext}
        \begin{equation}\label{eq_Jquant}
    J_C=2m\gamma\frac{1}{\lim\limits_{t\to\infty}\int_{-\infty}^{\infty} \! d\omega \, g(\omega) \, \int_0^t \! ds \, e^{-i\frac{\omega(t-s)}{m\gamma}-\frac{1}{\hbar}S'_{Im}[\theta_-^*;s,t]}}
\end{equation}
\end{widetext}
 In the limit of high temperature $\hbar\gamma\beta\ll 1$ and vanishing mass, the critical value \eqref{eq_Jclass} is recovered.
In this limit $K(\tau)\to\delta(\tau)$, and $\theta_-^*$ has non-zero value only on the time interval $[s,t]$, over which $\theta_-^*(t')=\frac{\hbar} {m\gamma}$ (see Appendix \ref{app:sp}).  This implies that $S'_{Im}[\theta_-^*]=\hbar \, \frac{k_BT}{m\gamma} \, (t-s)$, thus
\begin{eqnarray*}
J_C\biggl |_{\hbar\gamma \beta\ll 1} =
2 \left(\int_{-\infty}^{\infty} \! d\omega \, g(\omega) \,
\frac{k_BT}{(k_BT)^2+\omega^2}\right)^{-1}\!\!=J_C^{cl}(k_BT) \;.
\end{eqnarray*}
The last result corresponds to the classical result reported in the literature for $D=k_BT$ \cite{Acbron:review,Gupta:book}.
It is important to keep in mind that it holds only in the classical regime $\hbar\gamma\beta\gg1$, nonetheless we will extend this formula to low temperatures in order to provide a comparison with the behaviour of the quantum result in the following discussion.

\begin{figure}[H]
{
\begin{overpic}[width=.45\textwidth]{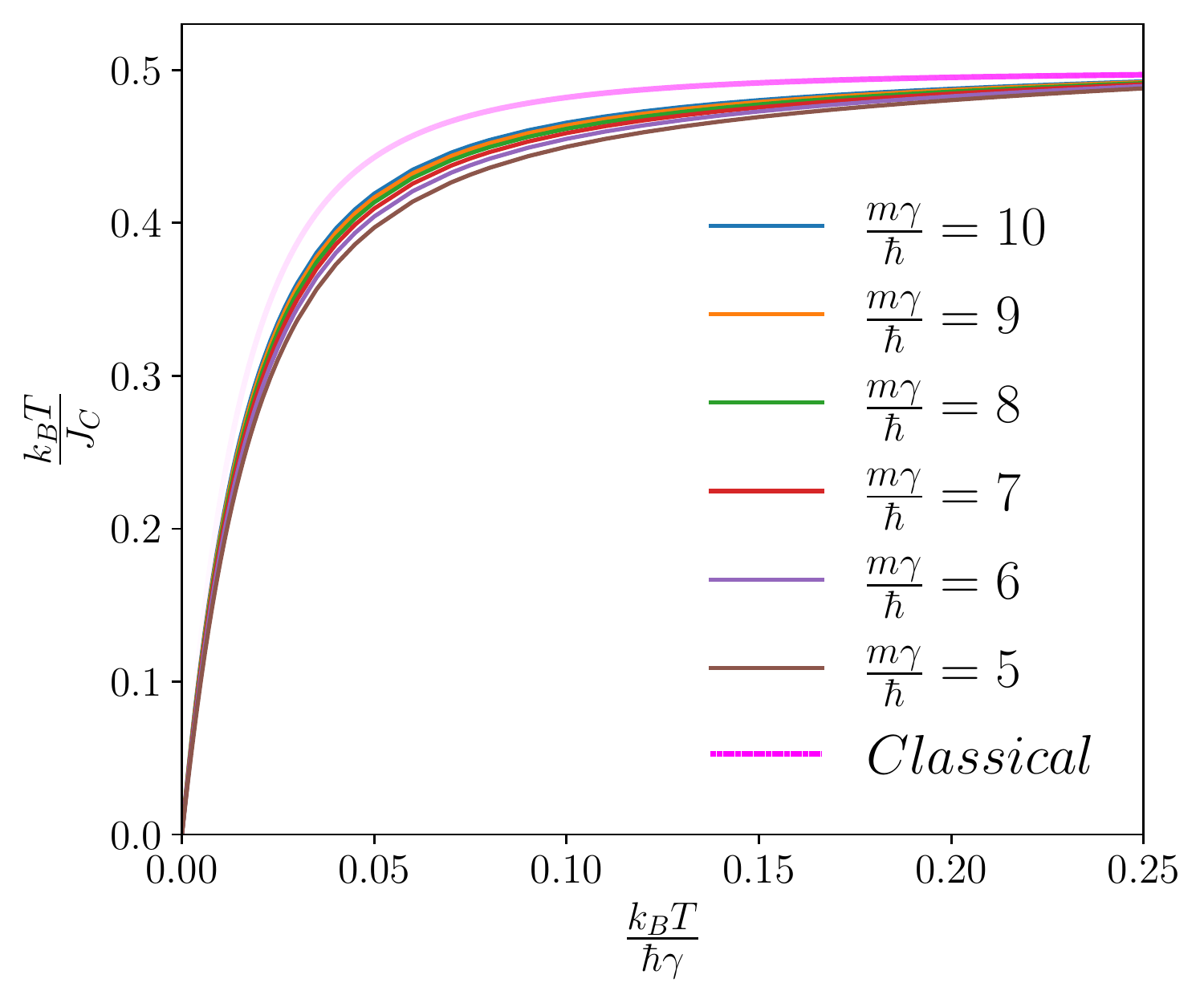}
\put (20, 71.5) {(a)}
\end{overpic}}\\
{\begin{overpic}[width=.45\textwidth]{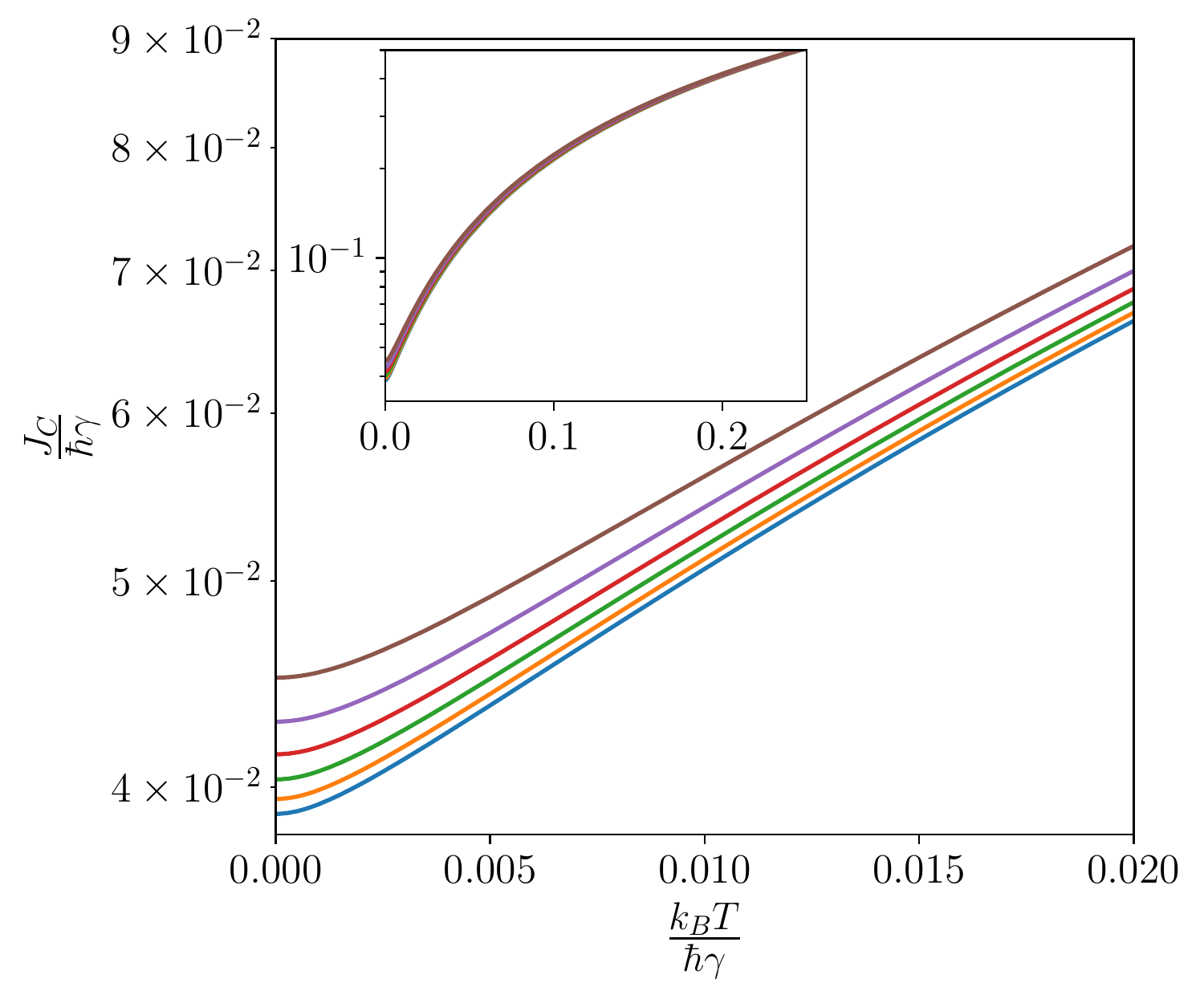}
\put (25, 71.5) {(b)}
\end{overpic}}\\
{\begin{overpic}[width=.45\textwidth]{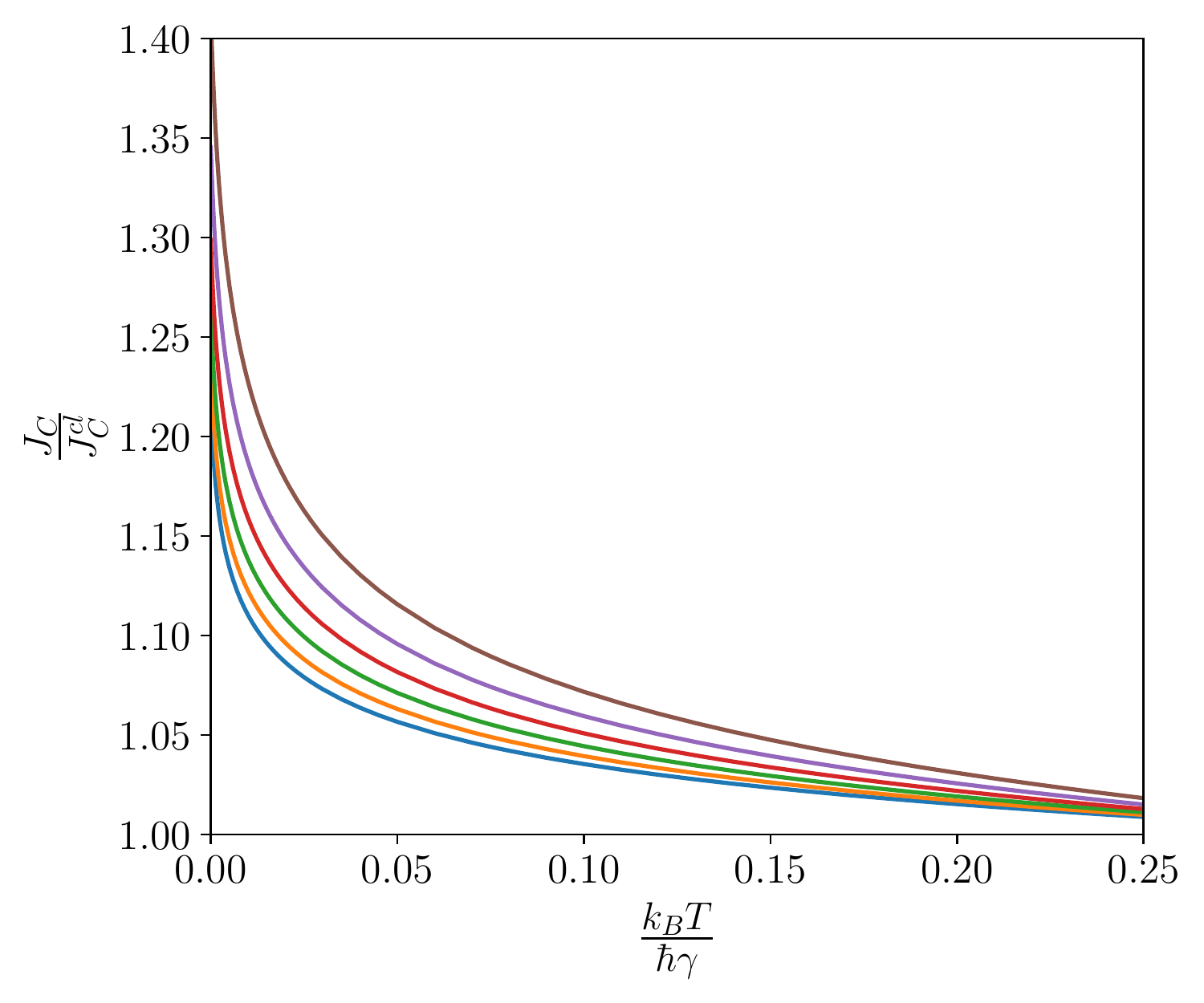}
\put (20,71.5) {(c)}
\end{overpic}}
\caption{Temperature-dependence of 
 the inverse of critical coupling in units of temperature (a) and (b) in logarithmic scale for different values of $\frac{m\gamma}{\hbar}$. The legend in (a) holds for all the panels. In (a) the classical critical coupling (pink line) is plotted for reference for high temperatures. (c) Ratio between the quantum results and the classical one (extended to low temperatures) vs temperature. The  difference in behaviour of the classical and quantum critical couplings is clearly evident  close to zero temperature. 
 All results are obtained for $g(\omega)$ Gaussian with zero mean and $\sigma=2$}
\label{fig1}
\end{figure}

The dependence of the critical coupling on temperature and the comparison with its classical counterpart are reported in
Fig. \ref{fig1}, where a Gaussian frequency distribution for the characteristic frequencies has been chosen to obtain the plots. From the analysis of panel (a) and (b),  the existence of three regimes emerges. In the \textit{classical regime}, defined by $\frac{k_BT}{\hbar\gamma}\gtrsim 1$, the classical results are recovered (cf pink line corresponding to the classical critical coupling for the overdamped massless model). The critical coupling in this latter case is \eqref{eq_Jclass}, and becomes $J_C^{cl}=2k_BT$ for $k_BT\gg\sigma$. The plots shows that this limit is reached asymptotically also by the quantum results.

A \textit{semiclassical region} is met when decreasing the temperature. Comparing the result in this region with the classical one extendend to lower temperature, we notice that the quantum results start to deviate quantitatively from the classical, although the behaviour of the quantum and classical critical coupling remains qualitatively similar. Notably, the quantum critical coupling is consistently higher than the classical one, as shown in panel (c). This is due to the emergence of quantum fluctuations, that as an extra source of noise, make the synchronization harder to be established \cite{Zueco:article}. The behaviour of $J_C$ in this region can be understood through the expansion in  $\theta_-\sim0$ already performed in Eq. \eqref{eq_classexp} to recover the classical limit. Going beyond the first order expansion therein, a semi-classical regime is  obtained from a third-order expansion\cite{Schmid:article}.\begin{widetext}
The latter yields a potential term:
\begin{eqnarray}\label{eq_deltaJ}
-Jr\sum_{s=\pm1}(-1)^s\cos\left(\theta_++s{\textstyle\frac{\theta_-}{2}}\right) \sim Jr\left(1+\Delta_J\right)\theta_-\sin\left(\theta_+ - \varphi\right) \;,
\end{eqnarray}
with $\Delta_J\propto \frac{\hbar}{m\gamma}\frac{\hbar\gamma}{k_B T}$ considering $\gamma$ to set the relevant time scale.
\end{widetext}
The above expansion tells us that in the semiclassical regime the motions happens in a potential of the same form of the classical one, but renormalized by $\Delta_J$. This yields the first deviation from the classical behaviour.

The \textit{quantum region}, characterised by $\frac{k_BT}{\hbar\gamma}\sim 0$ shows the most significant deviations from the classical result. In this region quantum fluctuations make the behaviour of the two critical couplings different and the deviation appears to be stronger than linear in the decrease of temperature

This is particularly evident from panel (c) showing the ratio between the quantum and the classical extended result $\frac{J_C}{J_C^{cl}}$. The ratio is greater than one, enforcing the fact the it is more difficult to reach the synchronized phase in the quantum regime, as it increases approaching zero temperature.

An important result emerging from this discussion is the existence of a finite critical coupling at any temperature ranging from $T=0$ to infinite $T$. A phase transition to a synchronized state is possible at every temperature and thus quantum fluctuations do not menage to prevent the emergence of this collective phenomenon.

\begin{figure}[h]
{\includegraphics[width=.45\textwidth]{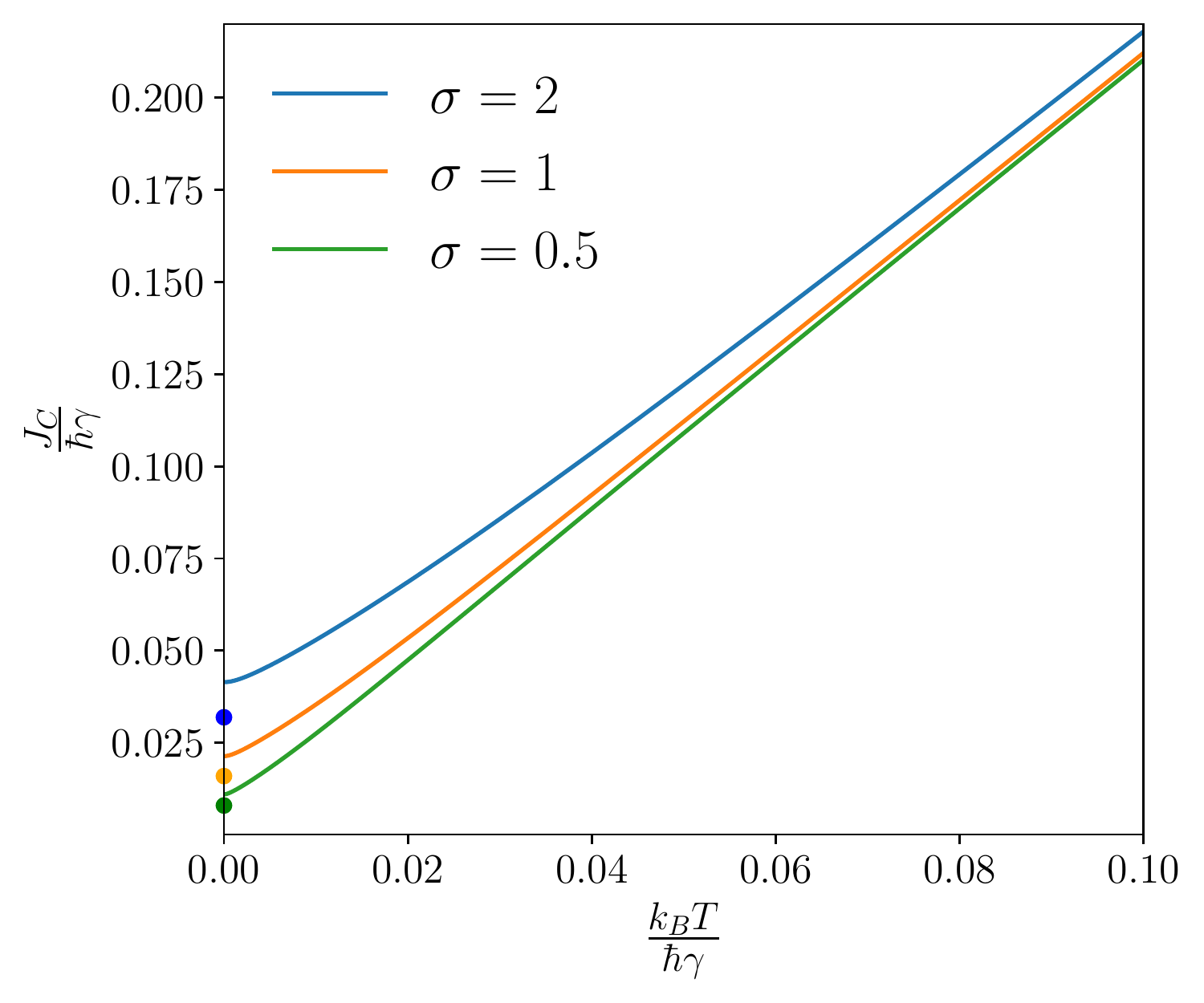}}
\caption{Critical coupling as a function of temperature for different choices of the variance $\sigma$ of the Gaussian frequency distribution. The overdamped ratio is fixed$\frac{m\gamma}{\hbar}=7$. The dots shown for $k_BT=0$ represent $\frac{J_C^{cl}}{\hbar\gamma}$ for the noiseless model compatible with $D=k_BT=0$. The colors of the dots are realated to the choice of $\sigma$ as shown in the legend.}
\label{fig2}
\end{figure}

Another analysis should be carried on. Figure \ref{fig2} shows the critical coupling for a fixed value of the overdamped ratio and for different choices of $\sigma$, the variance of the frequency distribution $g(\omega)$. This plot suggests that in the quantum realm, the width of the frequency distribution affects the critical coupling more than in the classical case. In both cases (quantum and classical), we notice that the wider the frequency distribution, the more difficult to synchronize. The quantum regime seems to be more affected by this effect.
This can be understood studying the behaviour of just two rotors. Suppose the rotors $\theta_1$ and $\theta_2$, have characteristic frequencies $\omega_1=\sigma$, $\omega_2=-\sigma$. Their phase difference $\Theta_-=\theta_1-\theta_2$ has a behaviour that is determined by the washboard potential $V[\Theta_-]=-\sigma\Theta_--J\cos(\Theta_-)$ and the coupling with the bath. If $\Theta_-$ is locked in a minimum of the potential, phase locking happens. From the shape (suppose $\sigma<J$) of the potential it is clear that decreasing $\sigma$, the height of the energy barrier that separates two minima ($\Delta V=2 J \sqrt{1-\frac{\sigma ^2}{J^2}}+2 \sigma  \sin ^{-1}\left(\frac{\sigma }{J}\right)-\pi  \sigma$) increases, making it easier to lock $\Theta_-$ in a minimum. Notice that, in the semiclassical region, Eq. \eqref{eq_deltaJ} suggests that the amplitude of the oscillations are given by $J(1+\Delta_J)>J$, explaining the enhancement of the disorder effect outside the classical regime,

It should also be noticed that for any finite variance $\sigma$ of the frequency distribution, the critical coupling at zero temperature is finite, yielding a quantum phase transition to a synchronized state.
The interplay of different element emerges from this plot: decreasing the the temperature increases the effects of quantum fluctuations and yields higher critical couplings with respect to the classical case, decreasing the width of the distribution helps the emergence of synchronization.

\begin{figure}[h]
{\includegraphics[width=.45\textwidth]{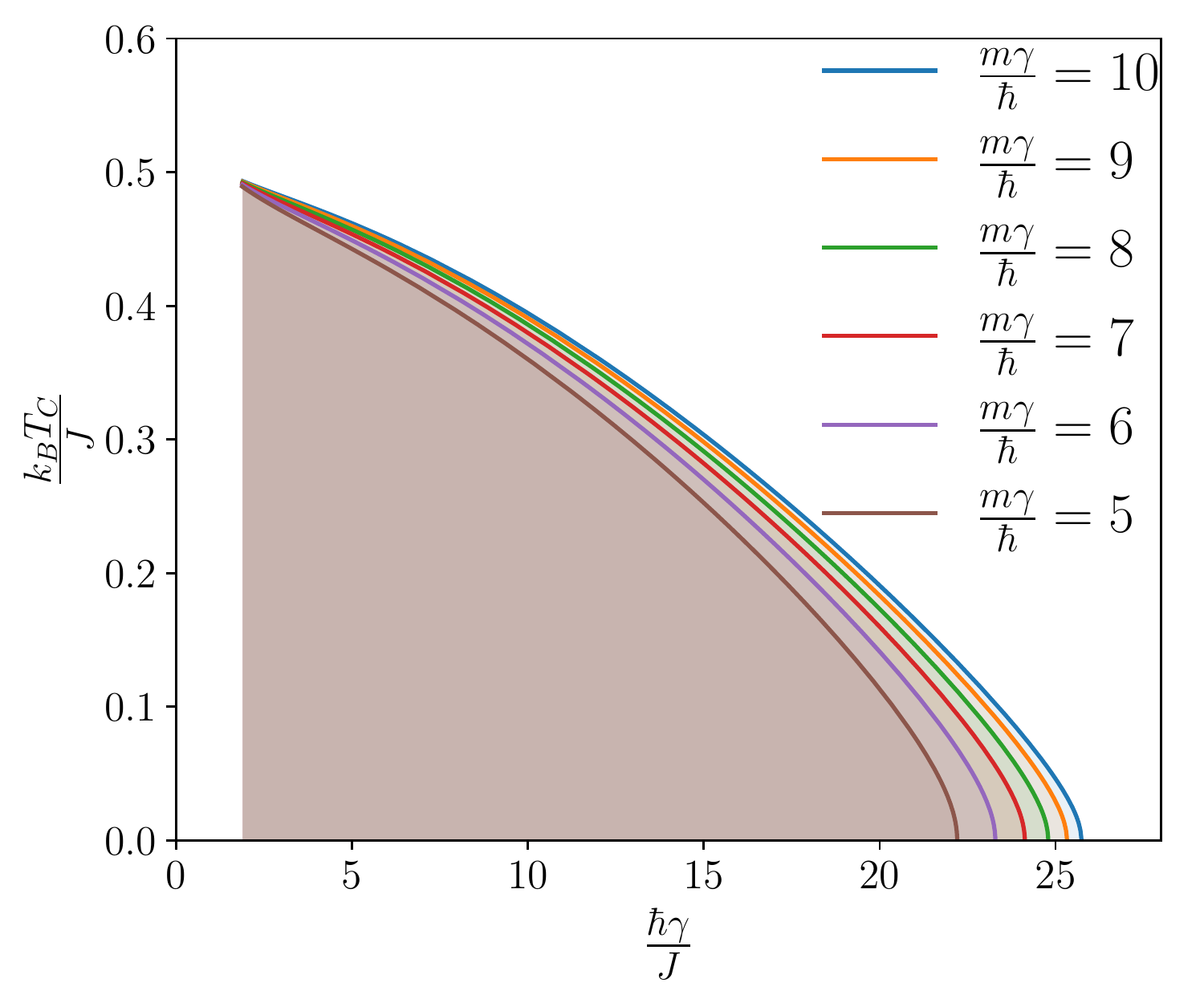}}
\caption{Phase diagram of the quantum model in the coupling-temperature space, where the coloured region indicates the synchronized phase.  A quantum phase transition at zero temperature is apparent. 
The result is obtained for a Gaussian distribution of frequencies with zero mean and variance $\sigma=2$.}
\label{fig3}
\end{figure}

These considerations are summarized in Figure \ref{fig3}, that shows the phase diagram of the quantum model. The perspective to study the model in this plot is reversed: the coupling is fixed to a value $J$ and the plot shows the behaviour of the critical temperature $T_C$, below which the system is synchronized. The synchronized phase corresponds to the coloured one in the plot, the white part corresponds to the incoherent motion phase. Two important features are captured in this picture. The quantum phase transition is evident: it is required a finite coupling to synchronize the system at zero temperature. Moreover, the classical result for the massless overdamped model, given by $k_BT_C=J/2$ is reached by our results in the high temperature limit.

\section{Discussion and conclusions}
\label{sec:conclusions}
In this work we introduced a generalization to the quantum regime of the well-known Kuramoto model. The model is built out of quantum interacting rotors coupled to environments modelled {\em \`a la} Caldera-Leggett as a collection of harmonic oscillators. The Feynman-Vernon technique allows us to obtain an evolution for the reduced density matrix that describes the subsystem of the rotors. The coherences in the reduced density matrix  are exponentially damped in the high-temperature limit, yielding a classical distribution that satisfies the Klein-Kramer's equation associated to the classical stochastic process that defines the noisy classical Kuramoto model.
The mean-field quantum model has been studied in its overdamped limit, that enables to perform a perturbative expansion around the critical coupling and carrying out the calculations analytically. 

This shows that the introduced quantum Kuramoto model in the overdamped regime admits a phase transition from a incoherent motion phase, to a synchronized one. The phase transition occurs at any temperature, yielding also a quantum phase transition at zero temperature. The critical coupling for the phase transition has been calculated analytically. It correctly reproduces the classical one in the high-temperature and shows deviation from this result extended to lower temperatures. In particular two regions can be observed, beyond the classical regime. A semiclassical region is first met when decreasing the temperature below the inverse damping rate $\frac{1}{\hbar\gamma}$, here quantum fluctuations make the critical coupling slowly deviate from the classical result, yet behaving qualitatively  similarly to the classical one. The quantum region, met around zero temperature $\frac{k_BT}{\hbar\gamma}\ll 1$, shows significant deviations from the classical result: around zero temperature a sudden increase of the ratio $\frac{J_C}{J_C^{cl}}$ happens, and at $T=0$ the critical coupling for the quantum model is higher than the classical one but finite (for finite variances of the distribution of the characteristic frequencies). The ratio between the quantum and the classical extended result is always greater than one, signaling the fact that quantum fluctuations play against the emergence of synchronization without destroying it.

\section*{Acknowledgments} 
We thank Stefano Ruffo for stimulating discussions. 
The research was partly supported by EU Horizon 2020 under ERC-ULTRADISS, Grant Agreement No. 834402.
R.F. and G.E.S. acknowledge that their research has been conducted within the framework of the Trieste Institute for Theoretical Quantum Technologies (TQT).
G.E.S and R.F. acknowledge financial support from PNRR MUR project PE0000023-NQSTI. 
G.E.S. acknowledges financial support from the project QuantERA II Programme STAQS project that has received funding from the European Union’s Horizon 2020 research and innovation programme under Grant Agreement No 101017733.
R.F. ackowledges financial support by the European Union (ERC, RAVE, 101053159). Views and opinions expressed are however those of the author(s) only and do not necessarily reflect those of the European Union or the European Research Council. Neither the European Union nor the granting authority can be held responsible for them.

\appendix
\begin{widetext}
\section{Path integral formulation for the classical stochastic process} \label{app:pi_class}
The average value over disorder of the observable $e^{i\theta(t;\omega,\xi)}$ must encode the fact that the phase $\theta$ satisfies the Langevin equation \eqref{eq_Lang}. With a path integral formalism and discretizing {\em à la Ito}, it gives:
\begin{eqnarray}\label{noise_av}
\expval{e^{i\theta(t;\omega;\xi)}}_\xi &=&
\N \int_{0}^{2\pi} \!\! d\theta
\int_{-\infty}^{\infty} \!\! dv \int \D \xi \int_{\theta(0)=\theta_0}^{\theta(t)=\theta} \D \theta \int_{v(0)=v_0}^{v(t)=v} \D v \,
\exp{-\frac{1}{4D}\int_0^tdt'\xi^2(t')+i\theta(t)} \nonumber \\
&& \hspace{35mm} 
\delta(\dot\theta_\tau-v_\tau) \,
\delta\big( m\dot v_\tau+m\gamma v_\tau-\force[\theta_\tau;\omega,\OP]+\xi_\tau \big) \;.
\end{eqnarray}
The second delta function, enforcing the Langevin dynamics, can be represented in exponential form through the use of an auxiliary field $\eta(\tau)$:
\begin{eqnarray}
\delta \big(m\dot v_\tau+m\gamma v_\tau-F[\theta_\tau;\omega,\psi]+\xi_\tau \big)) 
&\propto&
\int \D\eta \, \exp{i\int_0^t dt'\eta(t')\xi(t')} \nonumber \\ && \hspace{10mm}
\exp{i \int_0^t dt'\eta(t')\left[m\dot v(t')+m\gamma v(t')-\force[\theta(t');\omega,\OP(t')]\right]} \;.
\end{eqnarray}
Notice that choosing this representation of the delta function we now have a Gaussian path integral over the noise variable $\xi(t)$ with both quadratic and linear term. 
This integration generates another Gaussian form in the auxiliary variable $\eta$:
\begin{eqnarray}\label{eq_classaction}
\expval{e^{i\theta(t;\omega;\xi)}}_\xi &=&
\N \int_{0}^{2\pi} \!\! d\theta e^{i\theta}
\int_{-\infty}^{\infty} \!\! dv  
\int \D\eta 
\int_{\theta(0)=\theta_0}^{\theta(t)=\theta} \D \theta
\int_{v(0)=v_0}^{v(t)=v} \D v \, \delta(\dot\theta(\tau)-v(\tau)) \nonumber \\  
&& \hspace{10mm} 
\exp{-D\int_0^tdt'\eta^2(t')} \exp{i \int_0^t dt'\eta(t')\Big(m\dot v(t')+m\gamma v(t')-\force[\theta(t');\omega,\OP(t')]\Big)}
\;.
\end{eqnarray}
It is straightforward to see that the Gaussian integration over the auxiliary field $\eta$ yields:
\begin{eqnarray}
\expval{e^{i\theta(t;\omega;\xi)}}_\xi 
&=&
\N \int_{0}^{2\pi} \!\! d\theta \, e^{i\theta}
\int_{-\infty}^{\infty} \!\! dv  \int_{\theta(0)=\theta_0}^{\theta(t)=\theta} \D \theta \int_{v(0)=v_0}^{v(t)=v}\D v 
\, \delta(\dot\theta(\tau)-v(\tau))  \nonumber \\
&& \hspace{20mm}
\exp{-\frac{1}{4D}\int_0^t dt'\Big(m\dot v(t')+m\gamma v(t')-\force[\theta(t');\omega,\OP(t')]\Big)^2} \;.
\end{eqnarray}
The previous expression describes the average value of the observable $e^{i\theta(t;\omega,\xi)}$ for a stochastic process in the form of Eq.~\eqref{eq_Lang}. 
It is also convenient to write down the same average value for the same stochastic process in the massless limit. 
Keeping in mind that the only constraint is now satisfying the Langevin equation $\dot\theta=\force[\theta;\omega,\psi]+\xi(t)$, 
one can follow the previous steps and write:
\begin{eqnarray}
\expval{e^{i\theta(t;\omega;\xi)}}_\xi =
\N \int_{0}^{2\pi} \!\! d\theta \, e^{i\theta}
\int_{\theta(0)=\theta_0}^{\theta(t)=\theta} \D \theta
\exp{-\frac{1}{4D}\int_0^t dt'\Big(\dot \theta(t')-\force[\theta(t');\omega,\OP(t')] \Big)^2} \;.
\end{eqnarray}

\section{First order expansion of the self-consistent equation}
\label{app:firstorder}
The expansion to first order of the self-consistent equations along with the {\em Ansatz} $\varphi(t)=0$ and $r\sim 0$ constant, yields
\begin{eqnarray}
r &=& -\frac{iJ_Cr}{\hbar}\lim_{t\to\infty} \int_{-\infty}^{\infty} \! d\omega \, g(\omega) \, 
\int_{-\infty}^{\infty} \! d\theta_+ \, e^{i\theta_+} \, \int_{-\infty}^{\infty} \! d\theta'_{\pm} \,
\int_{\theta_+(0)=\theta'_+}^{\theta_+(t)=\theta_+} \! \D\theta_+
\int_{\theta_-(0)=\theta'_-}^{\theta_-(t)=0} \!\D\theta_-\nonumber\\
&&
\int_0^t \! ds \, \left(\cos(\theta_+(s)-\frac{\theta_-(s)}{2}) - \cos(\theta_+(s)+\frac{\theta_-(s)}{2})\right) \nonumber\\
&&\exp{\frac{i}{\hbar}\int_0^tdt'\left(m\dot\theta_-\dot\theta_+-m\gamma\theta_-\dot\theta_++\omega\theta_-\right)}
\exp{-\frac{D}{\hbar^2}\int_0^t dt'dt'' 
\theta_-(t') K(t'-t'') \theta_-(t'')} \,  \rho(\theta'_+,\theta_-') \;.
\end{eqnarray}
Rewriting the trigonometric term in exponential form, one gets straightforwardly
\begin{eqnarray}
r &=& -\frac{iJ_Cr}{2\hbar}\lim_{t\to\infty}\int_{-\infty}^{\infty} \! d\omega \, g(\omega) \, 
\int_{-\infty}^{\infty} \! d\theta_+ \, e^{i\theta_+}
\int_{-\infty}^{\infty} \! d\theta'_{\pm} \, 
\int_{\theta_+(0)=\theta'_+}^{\theta_+(t)=\theta_+} \! \D\theta_+ \,
\int_{\theta_-(0)=\theta'_-}^{\theta_-(t)=0} \!\D\theta_- \nonumber\\
&&\int_0^t ds \sum_{c,c'=\pm} c' \exp{ic\theta_+(s)-\frac{i cc'}{2} \, \theta_-(s)} \nonumber\\&&\exp{\frac{i}{\hbar}\int_0^t dt' \left(m \, \dot\theta_-\dot\theta_+ - m\gamma \, \theta_-\dot\theta_++\omega\theta_-\right)} \, 
\exp{-\frac{D}{\hbar^2}\int_0^t dt' dt''
\theta_-(t')K(t'-t'')\theta_-(t'')} \rho(\theta'_+,\theta_-') \;.
\end{eqnarray}
Denoting the argument of the exponential as $S'_{\eff}[\theta_+,\theta_-]$ and regarding it as a new effective action, one finally gets Eq.~\eqref{eq_effact}.

The path integration that now should be calculated is simply Gaussian. Decomposing the new effective action \eqref{eq_effact} in its real and imaginary part 
$S'_{\eff}=S'_{Re}[\theta_+,\theta_-]+iS'_{Im}[\theta_-]$, 
one can easily get the result of the path integration. The real part of the effective action contains quadratic and linear terms in both fields $\theta_+$ and $\theta_-$.
The imaginary part contains only a quadratic term in $\theta_-$. Notice that there are not quadratic terms just in the field $\theta_+$.

The strategy is now to find the saddle point of $S'_{Re}$, i.e. to find $\Tilde\theta_+(t')$ and $\Tilde\theta_-(t')$ such that the first derivative in both fields is zero, respecting the boundary conditions $\Tilde\theta_\pm(t)=\theta_\pm(t)$ and $\Tilde\theta_\pm(0)=\theta_\pm(0)$.

Then, the following change of variables should be performed: $\theta_{\pm}(t')=\Tilde\theta_{\pm}(t')+\delta\theta_\pm(t')$, with $\delta\theta_\pm(0)=\delta\theta_\pm(t)=0$. The path integration will be now performed over the fields $\delta\theta_+(\tau)$ and $\delta\theta_-(\tau)$. After having performed the change of variables, it is convenient to rewrite the action in matrix form:
\begin{eqnarray}
  &&S'_{\eff}[\delta\theta_+,\delta\theta_-;s,t]_{c,c'}=S'_{\eff}[\Tilde\theta_+,\Tilde\theta_-;s,t]_{c,c'}+\\&=&\frac{1}{2}\int_0^tdt'dt''\begin{pmatrix}\delta\theta_+(t')&\delta\theta_-(t')\end{pmatrix}A(t'-t'')\begin{pmatrix}\delta\theta_+(t'')\\\delta\theta_-(t'')\end{pmatrix}+\int_0^tdt'B(t')^T\begin{pmatrix}
      \delta\theta_+(t')\\\delta\theta_-(t')
  \end{pmatrix}\nonumber       
\end{eqnarray}
with
\begin{eqnarray*}
     A(t'-t'')&=&\begin{pmatrix}
      0&-m\delta''(t'-t'')+m\gamma\delta'(t'-t'')\\
      -m\delta''(t''-t')-m\delta'(t''-t')&\frac{2iD}{\hbar}K(t'-t'')\end{pmatrix}\;,\\
      B(t')&=&\frac{iD}{\hbar}\begin{pmatrix}
          0\\\int_0^tdt''\left(K(t'-t'')+K(t''-t')\Tilde\theta_-(t'')\right)
      \end{pmatrix}
\end{eqnarray*}
and $\N(t)=\frac{m\gamma}{2\pi\hbar(1-e^{-\gamma t})}$.
$A$ is the matrix that contains the coefficients of the quadratic terms in the fields $\Tilde\theta_-$, and $B$ contains the coefficients of the linear terms. Notice that the first entry of both $A$ and $B$ is zero, since there aren't quadratic and linear terms in the fiels $\delta\theta_+$ only.

The Gaussian path integration over the variables $\delta\theta_{\pm}$ yields
\begin{eqnarray}
\rho'(\theta_+,0,t) &=&
\frac{-i}{2\hbar}\sum_{c,c'=\pm1} c'
\int_0^t ds \int_{-\infty}^{\infty} \! d\theta'_{+} \, \int_{-\infty}^{\infty} \, d\theta'_{-} \, \N(t) \,
\exp{\frac{i}{\hbar}S'_{\eff}[\Tilde\theta_+,\Tilde\theta_-;s,t]_{c,c'}} \nonumber\\
&&\exp{-\frac{1}{4}\int_0^t dt' dt''
B(t')^T A^{-1}(t'-t'') B(t'')} \;.
\end{eqnarray}
Since the inverse of the matrix $A$ has the form $A^{-1}=\begin{pmatrix}
    a&b\\c&0
\end{pmatrix}$, the matrix product appearing above gives $B^TA^{-1}B=0$ and does not contribute to the result of the path integration.

Thus, the last issue we are left with is finding the solutions to the saddle point equations:
\begin{eqnarray}\label{eq_sp}
 \pdv{S'_{Re}}{\theta_+}\biggl|_{\Tilde\theta_\pm}&=&\ddot{\Tilde\theta}_-(t')-\gamma\dot{\Tilde\theta}_-(t')-\frac{\hbar c}{m}\delta(t'-s)=0\\
        \pdv{S'_{Re}}{\theta_-}\biggl|_{\Tilde\theta_\pm}&=&\ddot{\Tilde\theta}_+(t')+\gamma\dot{\Tilde\theta}_+(t')-\frac{\omega}{m}+\frac{\hbar cc'}{2m}\delta(t'-s)=0    \nonumber
\end{eqnarray}
The solutions to these equations have jump discontinuities in the derivatives, the jump is proportional to $\frac{\hbar}{m\gamma}$ thus it gets smaller in the overdamped limit. In the massless limit the discontinuities appear directly in the solutions (not just in the derivatives).
The solutions to the saddle point equations can be found in appendix \ref{app:sp}.

Substituting the solutions \eqref{eq_spsol1},\eqref{eq_spsol2}, in the first order approximation of the self-consistent equation, one gets in the infinite time limit

\section{Solutions to the saddle point equations} \label{app:sp}
The solutions to the saddle point equations \eqref{eq_sp} that respects the boundary conditions $\Tilde\theta_\pm(0)=\theta_\pm'$, $\Tilde\theta_-(t)=0$, $\Tilde\theta_+(t)=\theta_+$ are:

\begin{eqnarray}
\label{eq_spsol1}
 \Tilde\theta_-(t')&=&\begin{cases}
     \theta '_-+\frac{\left(e^{\gamma  t'}-1\right) \left(-\theta '_-+\frac{\hbar c  e^{\gamma  s}-\hbar c -\gamma  m \theta '_- e^{\gamma  s}}{\gamma  m \left(e^{\gamma  t}-1\right)}-\frac{e^{\gamma  t-\gamma  s} \left(\hbar c  e^{\gamma  s}-\hbar c -\gamma  m \theta '_- e^{\gamma  s}+\right)}{\gamma  m \left(e^{\gamma  t}-1\right)}\right)}{e^{\gamma  s}-1}&\hspace{2.6cm}0\le t'\le s \vspace{5mm}\\
     
     \frac{e^{-\gamma  s} \left(e^{\gamma  t'}-e^{\gamma  t}\right) \left(c \hbar e^{\gamma  s}-c \hbar-\gamma  m \theta '_- e^{\gamma  s}+\right)}{\gamma  m \left(e^{\gamma  t}-1\right)}&\hspace{2.6cm}s< t'\le t\\
 \end{cases}   \\
 \label{eq_spsol2}
 \Tilde\theta_+(t')&=&\begin{cases}
    -\frac{\left(e^{-\gamma  t'}-1\right) \left(\frac{e^{\gamma  t} \left(e^{-\gamma  s}-e^{-\gamma  t}\right) \left(-c \hbar c' e^{\gamma  s}+c \hbar c'-2 \gamma  m \theta '_++2 \gamma  \theta _+ m-2 t \omega \right)}{2 \gamma  m \left(e^{\gamma  t}-1\right)}+\theta '_+-\theta _++\frac{t \omega }{\gamma  m}\right)}{e^{-\gamma  s}-1}+\theta '_++\frac{\omega  t'}{\gamma  m}&\hspace{.2cm}0\le t' <s \vspace{5mm} \\
    
    -\frac{e^{\gamma  t} \left(e^{-\gamma  t'}-e^{-\gamma  t}\right) \left(-c \hbar c' e^{\gamma  s}+c \hbar c'-2 \gamma  m \theta '_++2 \gamma  \theta _+ m-2 t \omega \right)}{2 \gamma  m \left(e^{\gamma  t}-1\right)}+\theta _++\frac{\omega  t'}{\gamma  m}-\frac{t \omega }{\gamma  m}&\hspace{.2cm}s\le t'\le t\\
 \end{cases}   
\end{eqnarray}

It is interesting to notice that in the massless limit ($\frac{m\gamma}{\hbar}=$cost, $\gamma\to\infty$), the expressions of the saddle point equations get easier but discontinuities appear:
\begin{eqnarray}
\lim_{\substack{\gamma\to\infty\\m\gamma/\hbar=\rm{cost}}}\Tilde\theta_-(t')&=&\begin{cases}
        \theta'_-&\hspace{0.5cm}0\le t'<s\\
        \theta'_--c\frac{\hbar}{m\gamma}&\hspace{0.5cm}s<t'<t\\
        0&\hspace{0.5cm}t'=t
    \end{cases}\\
\lim_{\substack{\gamma\to\infty\\m\gamma/\hbar=\rm{cost}}}\Tilde\theta_+(t')&=&\begin{cases}
        \theta'_+&\hspace{0.1cm}t'=0\\
        \theta_++cc'\frac{\hbar}{2m\gamma}+\frac{\omega(t'-t)}{m\gamma}&\hspace{0.1cm}0<t'<s\\
        \theta_++\frac{\omega(t'-t)}{m\gamma}&\hspace{0.1cm}s\le t'\le t
    \end{cases}
\end{eqnarray}
\end{widetext}

\newpage
\bibliography{BiblioQSync}

\end{document}